# Intergenerational Mobility in China

Yi Fan, Junjian Yi, Junsen Zhang[1]

## Abstract

In this chapter, we provide a synthesis of empirical work on the intergenerational mobility in China, examining how much a child's success depends on their parents' success. We start with an overview of data sources and measure of intergenerational mobility in China's context, followed by discussion on the empirical findings in five categories of intergenerational mobility: income, education, social class, wealth, and health. Major channels of transmission of socioeconomic status across generations are investigated, including human and social capital, fertility and migration decision, housing price and trade liberalization in the market transition, and belief. Finally, we provide policy implications to promote upward mobility for children born to less advantaged parents in China's context.

---

[1] Yi Fan: Department of Real Estate, NUS Business School, National University of Singapore. yi.fan@nus.edu.sg. Junjian Yi: China Center of Economic Research, Peking University; National School of Development, Peking University. junjian.yi@gmail.com. Junsen Zhang: School of Economics, Zhejiang University. Department of Economics, Chinese University of Hong Kong. jszhang@cuhk.edu.hk. We thank Xiaoqian Zhang and Isabell Chew for excellent research support. Yi Fan acknowledges financial support from the National University of Singapore Academic Research Fund A-8002711-00-00. Junjian Yi acknowledges financial support from the National Natural Science Foundation of China (Grant No. 72595871, 72533001). Junsen Zhang acknowledges partial financial support from the Major Program of the National Social Science Fund of China (Grant No. 20&ZD076).

# Table of Contents

# 1. Introduction

Intergenerational mobility measures how much a child's success depends on his/her parents' success. The more mobile it is across generations, the more opportunities for children born to less advantaged parents to climb up the socioeconomic ladders. Van der Weide et al. (2024) compile a global education dataset from 400 household surveys in 153 countries (covering 97% of the world's population) and document that the intergenerational mobility is lower in developing countries compared to high-income ones. Figure 1 portrays the country-level expected rank of children in the 1980s cohort born to parents in the bottom 50% of education distribution in their generation. A higher value indicates greater intergenerational mobility.

As the largest developing country, China's intergenerational mobility in education is moderate. Using income as another measure of socioeconomic status (SES), Fan, Yi, and Zhang (2021) show that the intergenerational rank-rank correlation increases from 0.390 for the 1970–1980 birth cohort to 0.442 for 1981–1988 birth cohort, indicating declining intergenerational mobility. The magnitudes of intergenerational income and education mobility in China vary across gender, rural vs. urban areas, provinces, parents' SES, and family structure (Deng, Gustafsson, and Li, 2013; Zeng and Xie, 2014; Jin et al., 2019; Emran, Jiang, and Shilpi, 2020; Huang et al., 2021), implying complex dynamics in the transmission of SES across generations.

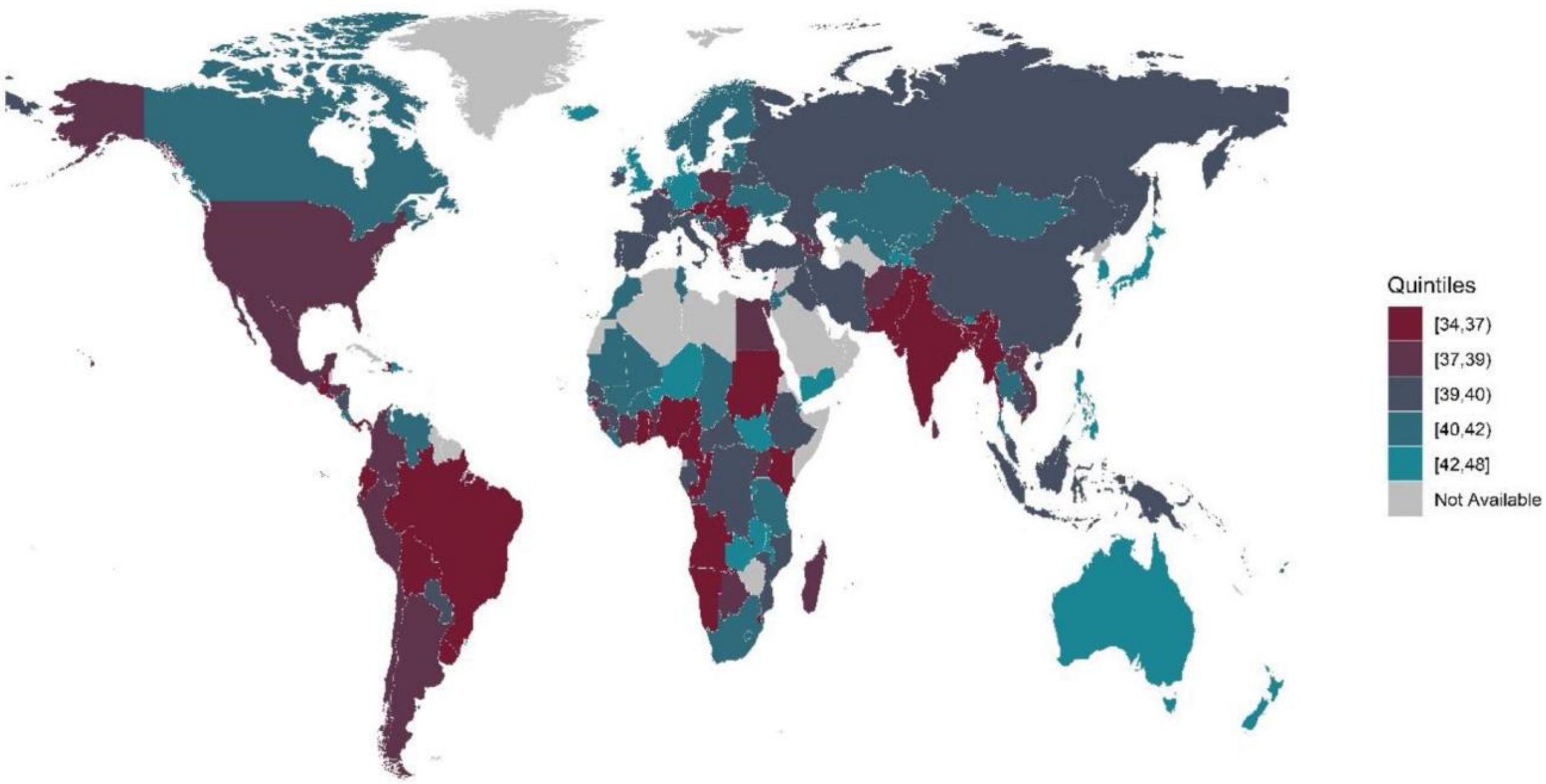

**Figure 1.** Expected education rank of a child in his/her generation born to parents from the bottom half of the education distribution from their generation.

Note: This is Figure 1 of Van der Weide et al. (2024).

In addition to the conventional measure of education and income as proxy of SES, the literature also estimates the intergenerational persistence across generations in social class, wealth, and health status in China. Specifically, occupation is a popular indicator for social class used by sociologists and economists. Emran and Sun (2015) show a significant rise in occupational mobility across generations in rural China from 1980s to early 2000, accompanying the relaxation of rural-to-urban migration. Nevertheless, the social fluidity across generations, which incorporates occupational mobility, declines with the deepening of China's market transition into the 2010s (Zhou and Xie, 2019), though the later anti-corruption campaign in 2012 reduces intergenerational occupation persistence by 18% (Ang et al., 2025).

Recent intergenerational literature extends from income to wealth, including broader assets such as real estate and savings. In China's institutional context of allocated public housing in the planned economy and escalating housing prices along with market reform of the housing market, the patterns of intergenerational co-residence and support change. Li and Wu (2019) show a causal impact of rising housing price on the increasing intergenerational co-residence in urban China from the 1990s into 2010s. Cui, Huang, and Wang (2020) and Zhu, Xin, and Chen (2022) find that parents who have benefited from the institutional housing advantages can help reproduce such advantages in their children's generation.

In term of mechanisms through which the parents' SES can be passed to the next generation in China, we summarize seven major channels: human capital, social capital, fertility and demographic structure, migration, housing prices, market transition and trade liberalization, and psychology. Consistent with classical literature and findings in the developed countries (Black and Devereux, 2011), education serves as the leading contributor to intergenerational mobility of SES. Specifically, it is evident that improved quality of primary and secondary education can promote upward mobility for children born to poor parents. However, the intergenerational persistence becomes stagnated in urban China after its higher education expansion (Liu and Wan, 2019; Huang, Huang, and Shui, 2021; Duan et al., 2022). Fan (2016) presents a shift in leading contributor to the intergenerational income persistence prior to and post China's market reform by household income category. In the post-reform era, the leading contributor among poorer families is education, while among richer households, it turns to social capital. Yuan and Chen (2013) find similarly significant contribution of social capital, which is generated by the principal component analysis on child's party membership, occupational type, and the ownership of working unit.

A unique institutional feature in China is about its population and mobility control. On the one hand, China imposes strict One-Child policy in urban areas from the late 1970s and allows rural families to have a second child if the first one is a girl from mid 1980s. The differential fertility between urban/richer households and the rural/poorer ones exacerbates income inequality across generations, making children of the rich stay rich, while children of the poor remain poor (Yu, Fan, and Yi, 2025). Gender bias towards girls are detects in this context, although it is found that girls in China benefit more from fathers' higher education than their counterparts from an analogy of India (Emran, Jiang and Shilpi, 2020). On the other hand, China strictly restricted rural to urban mobility within and across cities in the 1960s-1970s, as well as mobility across urban cities. The latter relaxation in mobility, along with the market reform, results in a large-scale internal migration and changes in the intergenerational mobility patterns. The findings are mixed though, as parents who benefit from migrating to richer areas can bring more financial resources to their children but scarifying time accompanying them (Sun, Huang, and Hong, 2012; Feng and He, 2022; Lei and Chae, 2024).

Since China's market reform in 1978, it has experienced a significant increase in the share of private enterprises, escalating housing prices, and accelerated trade liberalization. All those market factors have ramifications to the transition of SES across generations. For instance, the rising housing prices increases intergenerational co-residence rates. Parents who benefited from the institutionalized housing advantages are able to help with adult children's saving accumulation; thus the intergenerational status persists (Yuan and Lin, 2013; Rosenzweig and Zhang, 2019; Guo, Xia, and Zhang, 2022). The entry into the World Trade Organization (WTO) in 2001 also triggers change in intergenerational mobility, mainly through an income effect. Parents who benefit from the trade liberalization can accumulate more wealth and invest more in their children, promoting intergenerational education mobility at the prefecture level (Lou and Li, 2022). Recently, several studies examine beyond the conventional socioeconomic pathway and investigate a psychological channel. It shows that parents' mindset can affect outcomes of children, including personalities, attitudes, and cognitive skills (Roland and Yang, 2019; Huang, Song, and Xie, 2023).

Linking intergenerational income inequality with cross-sectional income inequality, a Great Gatsby Curve is presented in developed countries (Durlauf, Kourtellos, and Tan, 2022), showing a positive relationship between the two. The higher the cross-sectional inequality, the higher the inequality across generations is.

Fan, Yi, and Zhang (2021) and Zhang (2021) present a similar Great Gatsby Curve in China (Figure 2), indicating a transmission of income equality over time.

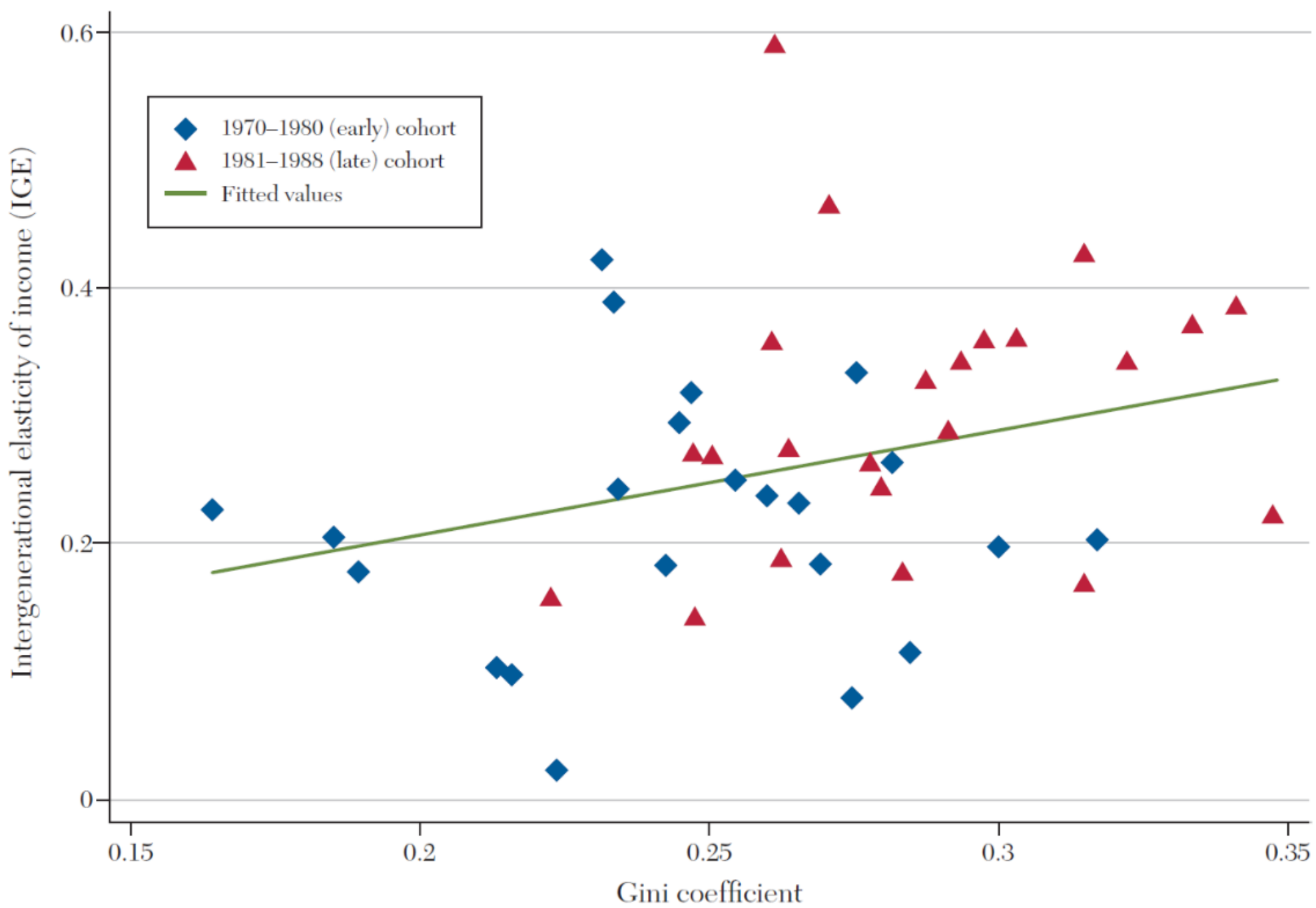

**Figure 2**. The Great Gatsby Curve in China using 25 provinces and municipalities by cohorts

Note: This is Figure 12 from Zhang (2021) which is reproduced from Fan, Yi, and Zhang (2021), Figure 5(a). The linear fitted line is estimated as IGE = 0.04 + 0.93Gini, where the estimated coefficient 0.93 is statistically significant at the 1% level.

To summarize, this chapter provides an overview of studies on the intergenerational mobility in China, in terms of theories, empirical estimates, and mechanism analysis. Different from developed countries which have administrative data or tax data to establish intergenerational linkage without co-residence bias and over long time period, China's intergenerational studies, most of which use household survey data, are subject to lifecycle bias, attenuation bias, and selection bias. With burgeoning big data and increasingly more high-quality surveys, the intergenerational studies in China can extend along various directions, such as linking with other socioeconomic datasets beyond families, expanding to multiple generations, and distinguishing between nature and nurture factors (Chen et al., 2023).

The rest of the chapter is organized as follows. Section 2 reviews the general models in intergenerational mobility and extension to China's context, as well as data and measurement challenges. Section 3 reviews empirical estimates in China's intergenerational mobility in income, education, social class, wealth, and health. Section 4 discusses transmission channels including human capital, social capital, fertility and migration decisions, housing price, market transmission, and mindset. Section 5 concludes with policy implications.

# 2. Theories, challenges, and data

## 2.1. Theories

Becker and Tomes (1979, 1986) pioneer in introducing families into the general economic analysis, defining the intergenerational mobility/persistence of income. Specifically, Becker and Tomes (1979) develop an equilibrium theory to understand inequality within and across generations. It identifies human capital investment by families, as well as the non-human capital factors (e.g., market factors), as determinants of individuals' income and quality of life. Their latter work (Becker and Tomes, 1986) supports the human capital transmission across generations and further details an economic model to characterize the intergenerational transmission of income, identifying a concave intergenerational income relationship under imperfect access to the capital market. They show that inequality in human capital tends to persist across generations and would only diminish after three generations. Solon (2004) rationalizes a log-linear regression to quantify the intergenerational income correlation, which is widely used in the empirical analysis of intergenerational mobility to estimate how sensitive a change in a child's income corresponds to that of parents' income.

Mogstad (2017) discusses the differences between Becker and Tomes (1979) and Becker and Tomes (1986), and introduces three significant extensions in application: multiple periods (e.g., crucial timing of investment during childhood), multiple skills (e.g., cognitive vs. non-cognitive skills), and various forms of investments (e.g., time inputs, monetary investments). Becker et al. (2018) further extend the earlier models of Becker and Tomes (1979, 1986), investigating whether intergenerational mobility results in cross-sectional inequality. The findings provide the first set of evidence on the Great Gatsby Curve that income inequality persists across generations. Parents with higher economic status invest more in their children compared to parents with lower economic status. Such variation in investment in children creates stagnation in social mobility, specifically affecting changes in individuals' economic status or social class. Cholli and Durlauf (2022) use linear and nonlinear regression models to measure the mobility, based on existing intergenerational mobility models (Becker and Tomes, 1979, 1986; Mogstad, 2017; Becker et al., 2018). Moreover, they summarize two sets of mobility mechanisms: family factor (income, education, credit constraints, household composition, and genes) and social factor (schools, neighborhood sorting, racial segregation, and peer and role model effects).

Specifically about China, Emran, Jiang, and Shilpi (2020) expand the intergenerational mobility models used in Becker and Tomes (1986) and Becker et al. (2018), by incorporating gender bias in the transmission of both human and non-human capital from parents to children. Their investigation in India and China shows that the father's education level significantly impacts the educational outcomes of daughters. Girls in China benefit more from their father's higher education than girls in India. In addition, Guo, Xia, and Zhang (2022) apply the Shapley-Shubik-Becker model to co-residence utility using the marriage surplus function in China's context. It analyses co-residence competition among multi-child families and between parents and parents-in-law, estimating determinants of co-residence decisions, such as gender, birth order, education, parental education, parent-child age gap, and local housing prices.

To sum up, the theoretical studies in the intergenerational mobility of China aligns with the classical models, though incorporating specific features prevalent in China's context to extend the model, such as gender bias and co-residence living.

## 2.2. Data and measurement challenges

Different from developed countries which usually have access to high-quality administrative data with little measurement errors, most studies of intergenerational mobility in China rely on survey data. Nevertheless, selection bias often arises from using the survey data. For instance, only family members living in the households or maintaining economic relationship with the households during the survey period are interviewed. Thus, there is co-residence bias if individuals are selective to stay at home. Married children who leave parents' households and establish their new ones are often not included in the survey. Similarly, temporary migrants, such as the rural-to-urban migrant workers, are not recorded or tracked in the household survey (Emran, Greene, and Shilpi, 2018; Fan, Yi, and Zhang, 2021).

To mitigate co-residence bias from imperfect data using truncated survey samples in China and broadly, developing countries, the literature endeavors with various remedies. Emran, Greene, and Shilpi (2018) show that the measure of intergenerational income correlation is more capable in reducing selection bias than intergenerational regression coefficients, to avoid truncated sample issue and selection bias in developing countries. Moreover, Shahe and Forhad (2019) find that the intergenerational rank correlation, as well sibling correlation, are more inclusive measurement of the intergenerational effects of parental education, family, and social factors. Deng, Gustafsson, and Li (2013), Fan (2016), and Fan, Yi, and Zhang (2021) adopt the Heckman selection model to address the selection bias in the context of estimating intergenerational mobility in China.

In addition to selection bias, other types of measurement challenges exist in the context of intergenerational mobility in China, with prominent examples of attenuation bias and lifecycle bias. Income in one survey wave could be subject to transitory shock in the specific year and thus may not be a proper measure of lifetime income. Solon (1989, 1992) show that the intergenerational mobility estimates could be downward biased in presence of the attenuation bias. Conventional method of averaging across different survey waves is used to mitigate the attenuation bias in China's and global context (Deng, Gustafsson, and Li, 2013; Yuan and Lin, 2013; Nybom and Stuhler, 2017), though Mazumder (2005) points out even averaging income across years could generate poor estimate for lifetime income if there is substantial persistence in transitory shocks. Some other studies use demographic or socioeconomic factors, such as education, as instrument for parental income in estimating the intergenerational mobility in China to partially overcome the attenuation bias (Gong, Leigh, and Meng, 2012).

Another frequently discussed measurement bias is lifecycle bias, as the association between current and lifetime earnings varies over life cycle. Specifically in the China's context, if children surveyed are systematically at early life stage, the estimate of intergenerational income persistence could be underestimated, as individuals with higher lifetime income may have steeper slope in the age-earning profiles. Gong, Leigh, and Meng (2012) document how such profiles differ across cohorts in China as influenced by the economic growth. To reduce the lifecycle bias, most studies restrict children's age to be in the mid-to-late life stage, which is evident to be least subject to lifecycle bias (Nybom and Stuhler, 2016). Fan, Yi, and Zhang (2021) compute lifetime income using Heckman model from survey data in China, aiming to mitigate co-residence, attenuation biases, and lifecycle bias.

Although with little access to administrative data in intergenerational studies in China, economists, sociologists, and other social scientists endeavor to construct high-quality survey datasets, to facilitate the exploration of intergenerational mobility and its changing patterns in China. Popularly used datasets include:

(1) China Family Panel Studies (CFPS): a longitudinal bi-annual household survey from 2010 to 2022, with details introduced in Xie and Hu (2015) and at https://www.isss.pku.edu.cn/cfps/index.htm;

(2) Chinese Household Income Projects (CHIP): a repeated cross-sectional household survey in 1988, 1995, 2002, 2013, and 2018, with details elaborated in Wan, Gustafsson, and Wang (2024) and at https://www.icpsr.umich.edu/web/ICPSR/series/243;

(3) China Health and Nutrition Survey (CHNS): a longitudinal household survey conducted every 2-4 years from 1989 to 2019. Details are introduced at https://www.cpc.unc.edu/projects/china;

(4) China Health and Retirement Longitudinal Study (CHRLS): a longitudinal bi-annual household survey from 2011 to 2020, with details presented at https://charls.charlsdata.com/index/en.html.

# 3. Estimates in intergenerational mobility in China

## 3.1. Intergenerational mobility in income

Intergenerational literature starts with theoretical interest in the persistence of income across generations (Becker and Tomes, 1979, 1986), and so does the following-up empirical research. Similarly in China, vast studies focus on investigating the magnitude of intergenerational association in income or earnings, as proxy for economic status. In this section, we review the estimate of intergenerational mobility in income in China, and its heterogeneity across time, region, and socio-demographic groups. We also link with the Great Gatsby Curve, to discuss the dynamics of inequality within and across generations.

Gong, Leigh, and Meng (2012) apply instrumental-variable (IV) strategy to address lifecycle bias and estimate that the father-son correlation in income is as high as 0.63 from the Urban Household Education and Employment Survey (UHEES) 2004 and the Urban Household Income and Expenditure Survey (UHIES) 1987-2004. Deng, Gustafsson, and Li (2013) use the CHIP data in 1995 and 2002 to estimate the intergenerational income elasticity (IGE) between fathers and sons in China, correcting for the co-residence bias with Heckman model. They show that the IGE increases from 0.47 to 0.53. Nevertheless, the intergenerational association between daughters and fathers is weaker. Taking direct transfer of human capital into account, Qin, Wang, and Zhuang (2016) estimate the IGE rises from 0.43 to 0.48 during the 1989 to 2009 period, with data derived from the CHNS. Fan (2016) shows a similar increase in the IGE from 0.43 to 0.51 for cohorts educated before and after China's market reform, using CHIP data in 1995 and 2002. For households with above average income in the post-reform era, the parent-child income persistence reaches as high as 0.71.

Regarding the heterogeneous effects, Yuan and Lin (2013) show the disparity in intergenerational income persistence between urban and rural China, using data from CHIP 1988-2002 and Chinese General Social Survey (CGSS) 2006. While both the IGE in urban and rural areas decline from 1988 to 2006, the urban estimate is persistently higher than the rural one. Yuan (2017) documents various degrees of intergenerational income mobility across income levels using longitudinal CHNS data 1989-2009. It is found poor families have relatively high mobility while wealthy fathers are likely to pass their favorable economic status to their sons. Using longitudinal CFPS data from 2010 to 2016, Fan, Yi, and Zhang (2021) impute the lifetime income for both parents and children and use various measure as proxy for the degree of intergenerational mobility, including IGE, intergenerational correlation, rank-rank estimate, and transition matrix of relative and absolute mobility. The IGE is estimated to increase from 0.39 (1970-1980 birth cohort) to 0.442 (1981-1988 birth cohort) and Figure 3 presents the percentage of children earning more than 100%, 120%, and 150% of their parents. It is found that the higher the parents' income, the lower the likelihood that children will surpass their parents economically. This decline is particularly pronounced among children whose earnings exceed 150% of their parents' income, and it is more evident for the late cohort than the early one (Panel C), reflecting China's rapid economic growth during the transition period. They also present that the decline in intergenerational income mobility is more apparent among urban and coastal residents, compared to their rural and inland counterparts.

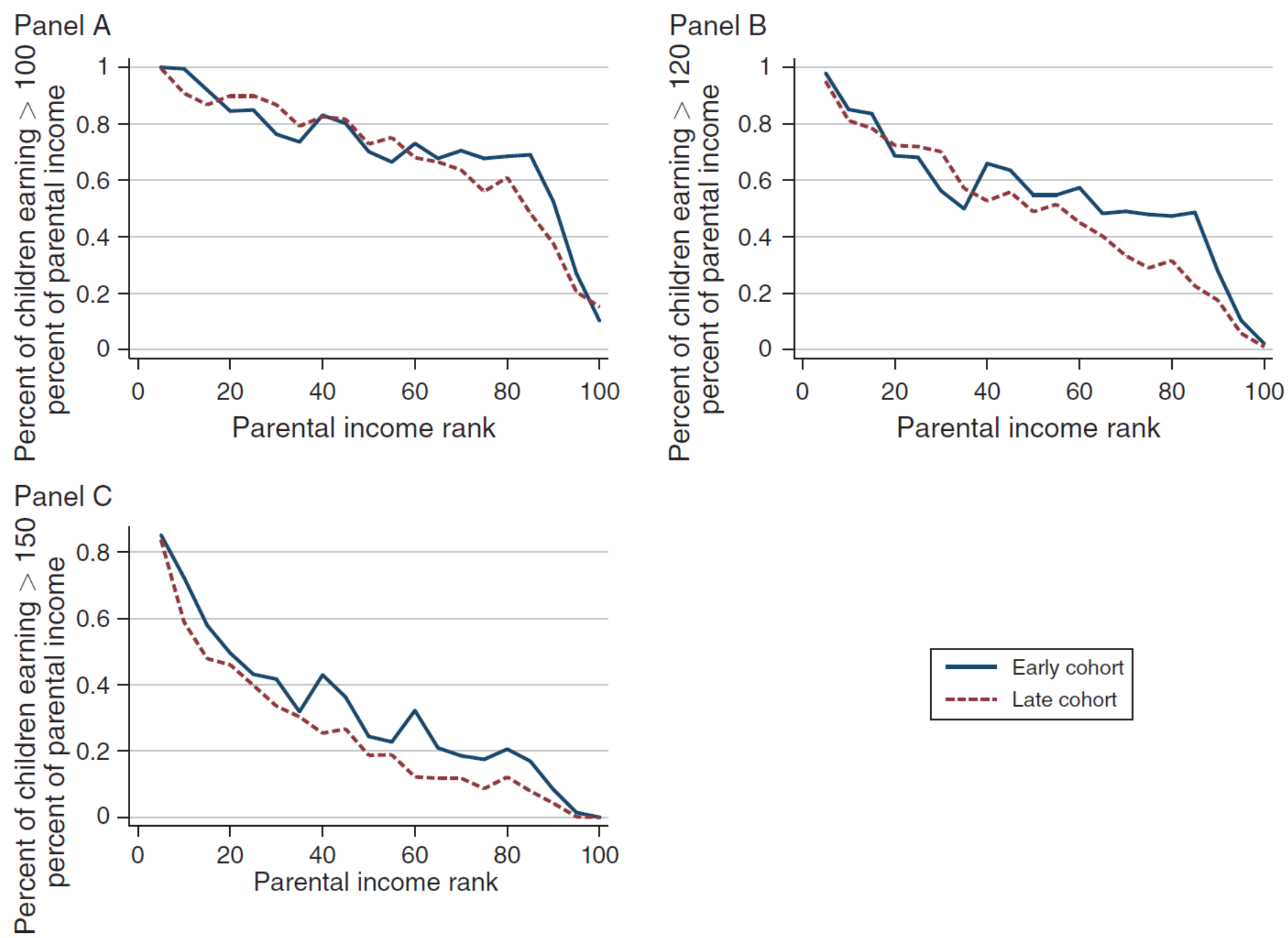

**Figure 3. Percentage of Children Earning More Than 100%, 120%, and 150% of Parental Income in 1970-1980 (Early) Cohort and the 1981-1988 (Late) Cohort**

Notes: This is Figure 4 in Fan, Yi, and Zhang (2021). Panels A, B, and C show the percentage of children earning more than 100 percent, 120 percent, and 150 percent, respectively, of parental income by cohort. Graphs are drawn by taking averages across five ranks of parental income in each cohort.

Table 1 summarizes data sources, study periods, and preferred estimates in the intergenerational income mobility research of China.

**Table 1. Summary of Intergenerational Income Mobility Estimates in China**

| Study | Data source | Time period / Cohort | Preferred estimates |
|---|---|---|---|
| Gong, Leigh, and Meng (2012) | Urban Education and Employment Survey (UHEES) 2004 & Urban Household Income and Expenditure Survey (UHIES) 1987-2004 | Working adults with father no older than 74 and mother no older than 69 in 2004 | Intergenerational income elasticities 0.634 for father-son; 0.973 for father-daughter;0.357 for mother-son;0.636 for mother-daughter |
| Deng, Gustafsson, and Li (2013) | CHIP 1995 & 2002 | Children aged 25-35 and co-resided with parents | IGE 0.47 (1995) and 0.53 (2002) for father-son pairs; 0.31 (1995) and 0.50 (2002) for mother-son; father-daughter pairs 0.40 (1995) and 0.33 (2022); mother-daughter pairs 0.25 (1995) and 0.45 (2022) |
| Yuan and Lin (2013) | CHIP 1988 to 2002, Chinese General Social Survey (CGSS) 2006 & CHNS 1989 to 2009 | Individuals aged 20 to 65, with less than a 10-year age difference between parent and child | Upward trend for rural IGE in 2005 & downward trend for urban IGE until 2005; Urban IGE with 0.51(1988), 0.42 (1995), 0.33 (2002), 0.30 (2006); Rural IGE with 0.42 (1988), 0.28 (1995), 0.22 (2002), 0.24 (2006) |
| Li, Liu, and Wang (2014) | CHNS 1989, 1991, 1993, 1997, 2000, 2004, 2006 and 2009 | Cohorts born on or after 1970 and before 1987 | Public education decreases intergenerational income immobility by 43%; IGE 0.83 |
| Chen (2016) | CHIP 1995 & 2002 | Working individuals in 1995 and 2002 | Income elasticity in China is 0.8 between 1990 and 1995; 0.4 between 1998 and 2002 |
| Qin, Wang, and Zhuang (2016) | CHNS 1989 to 2009 | Children above 25 years old and working | 0.052 increase in IGE when human capital is accounted |
| Yuan (2017) | CHNS 1989 to 2009 | Households with more than 16 years of survey records | IGE ranges 0.5 to 0.6 over the 1989 to 2009 period |
| Jin et al. (2019) | CHNS 1989 to 2011 | Father-son pairs aged between 20 and 65 years old | IGE 0.466 in 2011; minimum IGE of 0.225 in 2009; maximum IGE of 0.474 in 2004 |
| Fan, Yi, and Zhang (2021) | CFPS 2010 to 2016 | 1970-1988 birth cohorts | 0.390 for 1970–1980 birth cohort; 0.442 for 1981–1988 birth cohort |
| Huang et al. (2021) | CHNS 1989 to 2015 | Child cohorts born during 1973 to | IGE of 0.411 in low-spending provinces; 0.185 in high-spending |

| | | 1997 with father equals or less than 65 years old | provinces after controlling for birth years |
|---|---|---|---|
| Zhou and Bian (2024) | National Bureau of Statistics (NBS) 2010 and 1% 2005 | Random extraction of 20% of 1% 2005 census and 0.32% of 2010 census | 1 standard deviation increase in IGE, decreases net inflow rate by 0.017 between 2000 to 2010 |

### 3.2. Intergenerational mobility in education

One concern of using income, especially derived from household surveys, to measure intergenerational mobility is that it may not be a good proxy for lifetime socioeconomic status, as income is subject to transitory shocks in specific year, as discussed in Section 2.2. An alternative measure is education, which is more stable than income, especially after the mid-20s years old. In addition, there are less memory errors or misreporting in the educational years or levels. A bunch of literature has investigated the intergenerational mobility in education in China since the late 1990s.

Deng and Treiman (1997) use China's Cultural Revolution in 1966-1976, when many schools and universities were closed, as a quasi-natural experiment and estimate its impact on intergenerational persistence. They derive education data on father-son pairs from the 1982 census and find that the intergenerational persistence in educational attainment is weaker in China than other nations with comparable data, because of the social reshuffling during the Cultural Revolution. Meng and Zhao (2021) find a similar interruptive effect on children's educational attainment during the Cultural Revolution and the intergenerational effect is mainly through a reduction in parental education. Lou and Li (2022) use another economic shock, China's export expansion during 2000-2007 as a quasi-natural experiment, to estimate the change in intergenerational education mobility. Prefectures experiencing larger export shocks are more likely to have higher intergenerational education mobility, most likely due to an income effect.

Regarding the heterogeneous effects, Dong et al. (2019) examine the intergenerational education mobility in rural China, using the China Rural Development Survey (CRDS) in 2016. They find that the intergenerational transmission in education is significant for cohorts born after the 1980s but not for those born before the end of 1980. Emran, Jiang, Shilpi (2020) look into the gender inequality in the international education mobility and find that girls in China benefit more from their fathers' higher education than girls in a comparable counterpart of India.

In addition to measuring correlation between parents and children to estimate the degree of intergenerational persistence in educational attainment, the literature also expends to three generations and association among siblings. Zeng and Xie (2014) examine the effects of grandparents on children's schooling and show a significant influence of co-residing grandparents on dropout rates of grandchildren. Ahsan et al. (2024) analyse the intergenerational education mobility of 53 developing countries including China. They find that developing countries have a 44% higher sibling correlation compared to developed countries, indicating lower education mobility.

In sum, the intergenerational persistence in educational attainment in China is empirically evident, especially for younger cohorts born after 1980s. Comparing to an international analogy of India, girls in China benefit more from fathers' higher education. Institutional or economic shocks, such as the Cultural Revolution or the export expansion, do promote intergenerational mobility in education.

**Table 2. Summary of Intergenerational Education Mobility Estimates in China**

| Study | Data source | Time period / Cohort | Preferred estimates |
|---|---|---|---|
| Deng and Treiman (1997) | China Statistical Information and Consultancy Service Center 1989 on 1% public use sample in 1982 census | Cohorts born between 1945 and 1964 | Social origin effects return to normalcy for the 1957 and younger birth cohorts occurred after the Cultural Revolution; decline greater for intelligentsia sons comparing with cadre sons |
| Labart (2011) | CHNS 1991, 1993, 1997, 2000 & 2004 | Working individuals aged more than 16 years old | Increase in parents' education leads to a 0.20 year rise in children's schooling; increase in parents' income results in a 0.3946 year rise; having farmer parents leads to a 0.9463 year decrease |
| Golley & Kong (2013) | 2008 Rural–Urban Migration in China and Indonesia (RUMiCI) Survey | Children cohorts from 1941 to 1990 | 0.18 & 0.19 IGE for rural children; 0.21& 0.28 for urban children; 0.16 & 0.20 for migrant children |
| Li, Liu, Wang (2014) | CHNS 1989, 1991, 1993, 1997, 2000, 2004, 2006 & 2009 | Individuals born on or after 1970 and before 1987 | IGE of 0.830; a 43% higher in upward income mobility for those completed nine years of compulsory education compared to those did not |
| Zeng and Xie (2014) | CHIP 2002 | Rural children aged 7-18 and lived with both parents | -0.749 on dropout rates for children co-residence with grandparents |
| Magnani & Zhu (2015) | Chinese Population Censuses 1990 (1% sample) & 2000 (0.095% sample) | 5-year birth cohort aged between 21 and 25 | In 1990, intergenerational education transmission of father-son pair 0.249, mother-son 0.123, father-daughter 0.172, and mother-daughter 0.179; by 2000, they increased to 0.324, 0.202, 0.247, and 0.243 respectively |
| Mok (2016) | Youth survey conducted by author | College students or young adults between 18 to 34 years old | 54.1% of respondents perceived a decline in upward social mobility opportunities in Guangzhou; 77.9% in Hong Kong; 81.6% in Taipei |
| Dong et al. (2019) | China Rural Development Survey (CRDS) 2016 | Children above 16 years old | One year delay in mother's birth of children increase 0.077 to 0.106 years of children's schooling; Male children have on average 0.5 years more schooling years than females |
| Chen et al. (2019) | CHIP 2013 | Parent cohorts born between 1942 to 1966; Children cohorts born after 1961 | Intergenerational education persistence is 0.444 for urban children with a high school educated father; 0.281 for rural children; each lost year of father's schooling during Cultural Revolution cuts child's schooling by 0.596 (urban) and 0.540 (rural) |
| Guo, Song, Chen (2019) | CHIP 2013 | Children cohorts born in 1930 and later who completed their education | Urban children have an upward mobility rate of 48% with primary-educated parents and 68% with college-educated parents; Rural children have rates of 23% and 45% respectively; College expansion policy increases educational mobility by lowering the father's education impact by 0.069 in urban areas |
| Emran, Jiang, Shilpi (2020) | CFPS 2010 & 2016 | Children of the 18-35 age cohorts in 2016 | 0.09 intergenerational persistence for sons born to fathers without schooling; 0.35 for daughters |
| Meng and Zhao (2021) | CULS 2001, CHIP 2002 & Urban Residents Education and Employment Survey (UREES) | Parents born after 1940 and before 1963; Children born after 1964 and more than 20 years old | 18% reduction on children's probability to obtain university degree; reduction in education level by 0.32 years on children with Cultural Revolution interruptions of parental education |
| Lou and Li (2022) | 2010 survey | 2000 to 2007 cohort | Reduce in intergenerational education stagnation by 0.056 years |
| Xie et al. (2022) | Integrated Public Use Microdata Series (IPUMS) 1982, 1990, 2000, 2010, China One Percent Population Sample Survey 2005, Life Histories and Social Change in Contemporary China (LHSCCC) 1996 & | Birth cohorts between 1946 to 1985 | 0.24 intergenerational education correlation for 1946-1955 male birth cohorts; 0.40 for 1976-1985 male birth cohorts 0.36 for 1946-1955 female birth cohorts; 0.45 for 1976-1985 female birth cohorts |
| | CGSS 2005, 2006, 2008, 2010, 2011, 2012, 2013, 2015, 2017, 2018 | | |
| Zhao et al. (2023) | China One Percent Population Sample Survey 2015 & CFPS 2018 | Children cohorts aged 0 to 16 | Every 0.1 unit increase in intergenerational mobility, a decrease of 25.75% in average family education investment |
| Ahsan et al. (2023) | CFPS 2010 | Cohorts between 18-40 years old | 35.63% overestimation of relative intergenerational mobility for children from disadvantage families |

### 3.3. Intergenerational mobility in social class

In addition to the classical way of using income or education as measure for socioeconomic status, the literature also adopts other proxy for social status, such as occupation or employment status, in the context of intergenerational mobility in China.

Several sociological studies pioneer in this domain, such as Wu and Treiman (2007) and Walder and Hu (2009). Specifically, Wu and Treiman (2007) link China's hukou system with its intergenerational occupational mobility. They show that the distinctive population registration system fails to protect rural-origin males from downward mobility and only permits the best educated to obtain urban status. Walder and Hu (2009) examine the impact of revolution and reform on status inheritance in urban China, spanning 1949 to 1996. It is found that the Cultural Revolution policy effectively hindered the status transmission across generations for new elite families. Old elite families managed to retain some status transmission despite facing certain job position discrimination during the periods when these policies were imposed. Only the middle classes were able to transfer their elite status during the Mao period. Xie and Zhang (2019) support the finding on the downward mobility impact of the Cultural Revolution on children of the elite and upper-middle classes prior to the social reshuffling. They further expand the effect to the third generation and show that the intergenerational effect has faded due to the co-residing grandparent' effect.

Emran and Sun (2015) advance the understanding in intergenerational occupational mobility in rural China, comparing between 1988 and 2002 with CHIPs data. A significant increase in occupational mobility across generations is demonstrated, from agriculture to nonfarm occupations. Zhou and Xie (2019) harmonize six nationally representative surveys between 1996 and 2012 and show a decline in social fluidity across generations in China's market transition and industrialization. The social fluidity is derived from the dimensions of status hierarchy, class immobility, and affinity. Nevertheless, comparing with other 11 advanced industrial countries, the social fluidity in China is still considered high by international standards. Recent literature extends to the multi-generational effects. Mare and Song (2023) use Genealogical data from the Qing Dynasty Imperial Lineage and from population registry data for Liaoning province to show that the multigenerational influence is more multifaceted than earlier speculations. It can have profound impacts on marriage, fertility and status attainment over longer term. Ang, Qin, Tan and Zhang (2025) use China's 2012 anti-corruption campaign as a quasi-natural experiment and show that college graduates with cadre parents are more likely to work as civil servants. However, such occupational persistence is reduced by 18% after the anti-corruption campaign.

In sum, most literature in this stream, with a majority from the sociological perspective, complements the classical method of using income or education as measure of socioeconomic status. It generally shows an increase in intergenerational mobility in social class for cohorts born in China's revolution period, due to the social reshuffling. However, a decline in mobility is observed among cohorts born during the market-oriented reform era, as the emergence of markets provides opportunities for the privileged classes to pass on socioeconomic advantages to their children.

**Table 3. Summary of Intergenerational Mobility in Social Class in China**

| Study | Data source | Time period / Cohort | Preferred estimates |
|---|---|---|---|
| Wu and Treiman (2007) | Life Histories and Social Change in Contemporary China Survey 1996 | Male sample between 20-55 years old | 0.703 immobility for nonagricultural workers; 1.9797 agricultural workers (farmers); upward social mobility from education is 0.250 |
| Walder and Hu (2009) | Life-history survey of 3,087 adults | Urban sample aged 20-69 | Offspring of old elite were 83% more likely to retain elite status during the Mao era; 49% probability to maintain intergenerational elite status after control for parental education |
| Emran and Sun (2015) | CHIP 1988 & 2002 | Adults aged between 18-60 | 0.09 for daughters with parents work in non-farm participate in non-farm sector in 1988; 0.33 with at least one parents in non-farm sector; 0.71 with non-farm status parents; 0.43 with parents in agriculture in 2002; 0.73 with parents in non-farm sector; similar trend for sons |
| Zhou and Xie (2019) | Life Histories and Social Change in Contemporary China (LHSCCC) 1996 & CGSS 2005, 2006, 2008, 2010 and 2012 | Cohorts aged 31 to 64 and born after 1936 | 7.95 estimated coefficient of social economic status for men born in 1980;10.45 for women |
| Xie and Zhang (2019) | CFPS 2010, 2012, 2014, 2016 | Cohorts born between 1940 and 1969 | Net downward educational mobility rate of 17% for child from elite classes; 10% from upper-middle classes; 5% upward class mobility rate from the grandparent generation to the child generation (family with poor peasant class); 12% for worker class and red class |
| Mare and Song (2023) | China Multigenerational Panel Dataset Imperial Lineage (CMGPD Imperial Lineage) and the China Multigenerational Panel Dataset-Liaoning (CMGPD-LN) | Residents and their descendants in east Liaoning (1749-1909) and in Qing Dynasty (1636-1912) | Qing Dynasty: 2.9 times as many high-status descendants as counterparts with father and grandfather effects; 3.2 with inclusion of great-grandfather effects; Liaoning: 2.8 and 4.7 respectively |
| Ang, Qin, Tan and Zhang (2025) | Chinese College Students Biannual Employment Survey (CCSES) 2007-2019, China's Corruptions Investigations Dataset (CCID) and Procuratorial Yearbook of China, National Tax Survey Database (NTSD) | Employed Associate or Bachelor graduates | When fathers hold managerial roles in government institutions, their children are 60.1% more likely to enter the civil service upon graduation, as compared to those whose fathers are in other occupations; China's 2012 anti-corruption campaign reduces occupational persistence by 18% |

### 3.4. Intergenerational mobility in wealth

In addition to examining the persistence in income across generations, intergenerational studies are also interested in the transmission of wealth, which includes broader categories of assets such as property and savings, from one generation to the next. Due to the Confucius tradition in China, passing on wealth, especially housing assets, to the offsprings is a classical way to maintain social class. Thus, a bunch of literature investigates the interactive effects between housing and intergenerational mobility in China.

Li and Shin (2013) study the changing pattern of intergenerational housing support between retired old parents and children and identify the effect of allocated public housing in the flow of intergenerational support. Li and Wu (2019) link rising housing price with the increasing intergenerational co-residence in urban China from the 1990s into 2010s. Such causal impact is more evident among those who are not homeowners, with a lower level of wealth, and relatively younger. Cui, Huang, and Wang (2020) analyze the role of family origin in reproducing housing inequality for the young generation. It is found that the housing advantages obtained from the institutional benefits during the housing reform era are transferred to their offspring, and vice versa.

You et al. (2021) and Zhu, Xin, and Chen (2022) investigate the intergenerational effects between housing prices and education. Using the China Household Finance Survey, You et al. (2021) show concave slopes of children's education as a function of father's education, as a result of tightening household credit constraints due to the high housing prices. Zhu, Xin, and Chen (2022) use the 2006 Chinese General Social Survey (CGSS), showing that political elites' housing advantage can be retained in their children's generation.

Due to data limitations in tracking housing status or transaction prices across different generations, current literature in China's context cannot explicitly quantify the intergenerational association in housing wealth.

Table 4 summarizes major findings from existing research and future endeavors are warranted when there are abundant data on housing transaction records linking across generations.

**Table 4. Summary of Intergenerational Wealth Mobility Estimates in China**

| Study | Data source | Time period / Cohort | Preferred estimates |
|---|---|---|---|
| Li and Shin (2013) | Collected data from survey in February and April 2009 | Interviewees from families with at least one retiree (65 years and above) and have at least one child | Parents who are allocated public housing are less likely to receive housing support from their children; those parents unlikely offer more support to their children than parents who were note allocated public housing |
| Li and Wu (2019) | CHNS 1991 to 2011 & 2005 Inter-Census Population Survey | Urban married adults aged between 20 and 50 | with 1% increase in housing price, there is 0.317 percentage point increase of co-residence rate with elderly parents/parents-in-law |
| Cui, Huang, and Wang (2020) | 2013 Fudan Yangtze River Delta Social Transformation Survey | Cohorts born between 1980-1989 | -0.339 homeownership transmission for men whose parents do not own a home compared to women; 0.141 for men whose parents own a home compared to women (0.090) |
| You et al. (2021) | CHFS 2011 | Individuals born after 1976 and aged 20 or below | 0.098 for average maternal IGE coefficient; 0.375 for paternal association when account for house prices |
| Zhu, Xin, and Chen (2022) | CGSS 2006 | Respondents aged 18-65 | 2% increase in child's housing assets with every additional year of father's education; 50,245 RMB premium on child's housing assets from fathers' Party membership; more than 50,000 RMB premium for those with fathers worked in state sector |

### 3.5. Intergenerational mobility in health

In recent decades, a stream of burgeoning literature investigates the intergenerational mobility in health status and health behaviors in China. Using the CHNS 1991-2009 data, Eriksson, Pan, and Qin (2014) demonstrate a strong correlation of health status between parents and children in both urban and rural China. As high as 15%-27% of the rural-urban inequality of child health can be attributable to inequality endowed from parents' health. Pan and Han (2017) advance the understanding in the intergenerational persistence of health behavior using the same datasets. They find that parents' smoking behavior is positively associated with children's decision to take up smoking and the amount of cigarettes to consume. Adolescents with both parents smoking have 4.8% higher probability of smoking, compared to counterparts with nonsmoker parents. Adopting administrative health records from Taiwan, Chang et al. (2023) present a rank-rank slope of 0.22 in health between adult children and their parents. The intergenerational persistence in health is more pronounced in maternal transmission, among parents with sons, and at the upper tail of the parental health distribution. However, it appears to be unrelated to household income levels and is not primarily driven by genetic factors.

**Table 5. Summary of Intergenerational Health Mobility Estimates in China**

| Study | Data source | Time period / Cohort | Preferred estimates |
|---|---|---|---|
| Eriksson, Pan, and Qin (2014) | CHNS 1991 to 2009 | Children less than 18 years old | 15%-27% of rural-urban inequality in health transmission |
| Pan and Han (2017) | CHNS 1991 to 2009 | Adolescents aged between 13 and 18 | 4.8% increase in the probability of smoking habit in adolescents with smoking parents compared to those with nonsmoking parents |
| Kong, Osberg, and Zhou (2019) | CHNS 1997, 2000 and 2004 | Children between 4-18 years old | 1.8 kg gain and 2.2 percentage-point increase in the overweight rate of medium-build 10-year-old boys |
| Chang et al. (2023) | Taiwan's National Health Insurance (NHI) system 2000-2019 | Cohorts born between 1979 and 1981 | 0.22 rank-rank slope in health; Intergenerational health association level is 0.26 |

# 4. Mechanisms for intergenerational mobility in China

Figure 4 illustrates the channels of intergenerational transmission in SES discussed in China's literature. Appendix A1 summarizes details of findings in those papers of intergenerational transmission mechanisms.

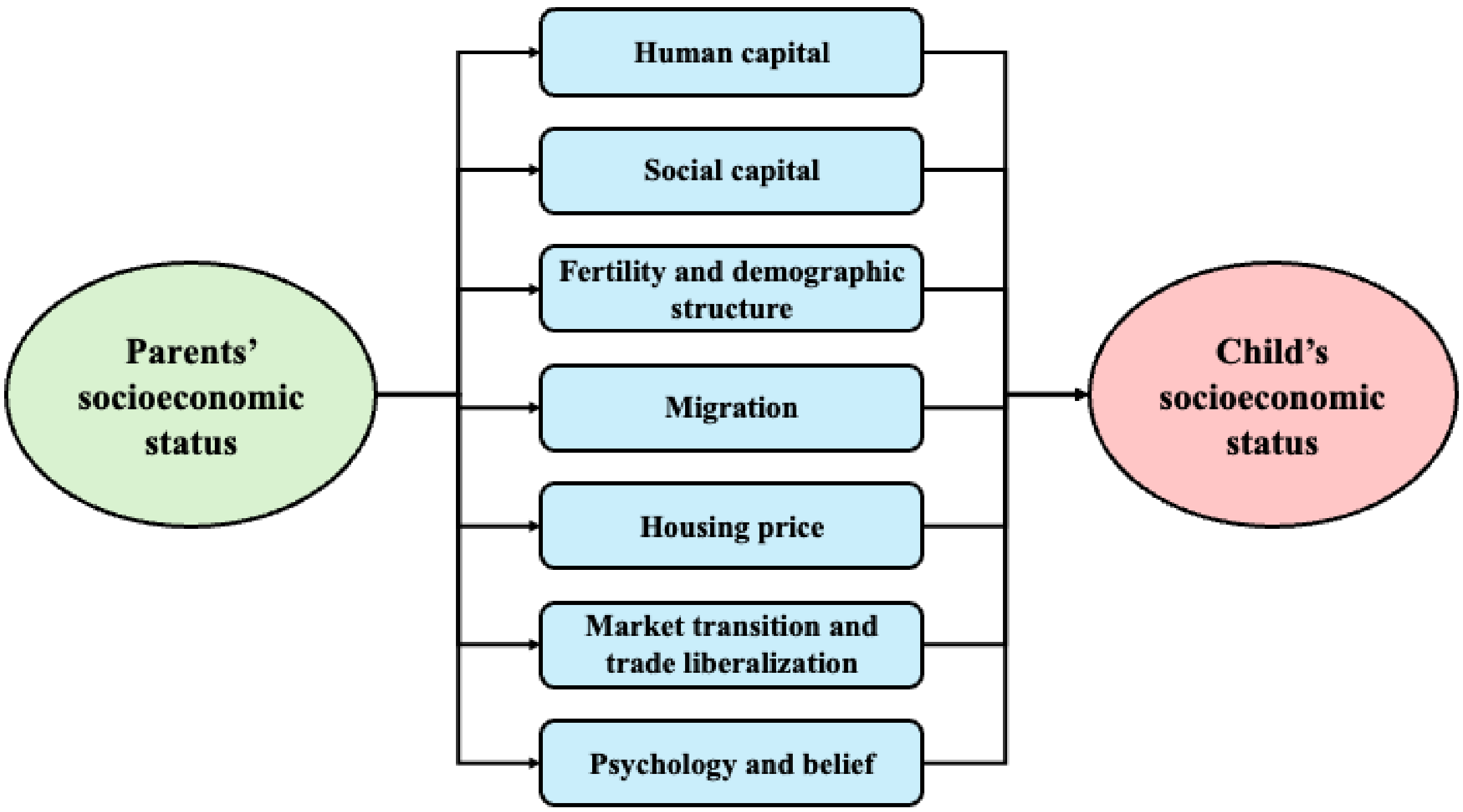

**Figure 4. An Illustration of Intergenerational Transmission Channels in China**

## 4.1. Human capital

The most studied, and possibly most prominent channel for intergenerational transmission of socioeconomic status (SES) is human capital accumulation, specifically education. With less stringent budget constraints and probably preference for higher education, parents with higher socioeconomic status tend to invest more in children's human capital. As a key determinant of earnings, the higher human capital can translate into higher income, indicating persistence in socioeconomic status across generations.

In China, because of its historically meritocratic educational institutions, tracing back to the "Keju" (Chen, Kung, and Ma, 2020) and inherited by the National College Entry Examination (Gao Kao), human capital accumulation is an important way, if not the most, to achieve upward social mobility. Chen et al. (2015) present a robust U-shaped pattern in the intergenerational persistence in educational attainment for urban Chinese born between 1930 and 1985. The intergenerational persistence tends to rise among cohorts educated in the market reform era and likely correlates with changes in China's political, economic, and educational institutions.

Qin, Wang, and Zhuang (2016) and Yang and Qiu (2016) use three- and four- period overlapping-generation models, respectively, supporting the remarkable role of human capital transfer, especially investment in early education, in determining intergenerational persistence in SES. Using the plausibly exogeneous surge of China's higher education expansion as a quasi-natural experiment, Liu and Wan (2019) show that the education expansion reduces intergenerational transmission of education, but when taking the quality of higher education into account, such significant effect disappears.

Huang, Huang, and Shui (2021) and Duan et al. (2022) further study the effects of primary & secondary and tertiary education on intergenerational socioeconomic mobility, showing that better quality of primary and secondary education under government education expansion benefits poorer families more and thus promote intergenerational mobility. Nevertheless, the higher education appears to have changed from promoting intergenerational mobility to stagnating intergenerational persistence in urban areas after China's higher education expansion. Zhang and Gu (2025) also show that while China's college expansion increases average human capital and income, it reduces mobility by making college attendance more dependent on family background. They echo earlier literature (Chen et al., 2015; Liu and Wan, 2019) on the inequality of access to higher education after the higher education expansion.

Kong, Osberg, and Zhou (2019) investigate other less studied human capital – health. Using the massive layoffs during China's State-Owned Enterprise (SOE) reform as a quasi-natural experiment, they demonstrate causal impact of 1.8 kg weight gain and 2.2-percentage-point increase in the overweight rate due to the reform. Parents' anxiety in economic insecurity can be translated into inferior health status of children.

In sum, consistently with international studies, the literature on China shows a significant role of human capital (education and health) in intergenerational transmission of SES. Nevertheless, political, economic, and educational institutions in different regimes have differentiated effects, signaling the importance of analysis by different contexts of geographic regions, policy era, and time periods.

### 4.2. Social capital

Beyond classical channel of human capital, another stream of the literature investigates the role of social capital in the intergenerational transmission of SES in China. The social capital is usually composed of occupation, industry and ownership of the work unit, and/or party affiliation. Specifically, Zhang and Eriksson (2010) investigate the inequality of opportunity in China and find that parental income and their type of employment category (farming, collectives, private enterprises, government or state-owned enterprise (SOE), foreign owned enterprise) can explain as high as two thirds of the total inequality of opportunity. Interestingly, parental education only plays a minor role, implying significant contribution from the parental connections. Yuan and Chen (2013) find similar significant contribution of social capital, which is generated by a principal component analysis on child's party membership, occupational type, and the ownership of working unit. Moreover, they show that wealth contributes the most compared to human capital and social capital. The three channels in total can explain more than 60% of variations in the intergenerational income mobility in China.

Fan (2016) demonstrates a shift in leading contributor to the intergenerational income persistence prior to and post- reform era by household income category. For below-average income households in urban China, the leading contributor shifts from ownership of work unit to education across market reform. However, for above-average income households, the leading contributor remains social capital and has an increasing share among the three competing channels. Emran et al. (2023) extend to rural China and find fathers' nonfarm occupation and education are separable in shaping intergenerational educational persistence in China. Interesting, genetic correlations alone can largely explain for the persistence across generations with data free of co-residence bias. Comparing to a geo-economic analogy of India which has contrasting evidence, the farm-nonfarm differences in return to education and geographic mobility are plausible reasons.

### 4.3. Fertility and demographic structure

In addition to human and social capital, recent studies investigate the role of fertility decision, as well as family's demographic structure, in the transmission of economic status across generations in China. Using the staggered rollout of the One-Child Policy (OCP) across province and birth cohort as a quasi-natural experiment, Yu, Fan, and Yi (2025) show that differential fertility between rural/poorer and urban/richer households exacerbates intergenerational income inequality. The rural families are less constrained by the OCP than their urban counterparts and tend to have more children but invest less in each child's human capital. Thus children of the urban/wealthier parents stay rich, decreasing the intergenerational income mobility. The back-of-the-envelope calculation suggests that approximately ¼ of the decline in the intergenerational income mobility in China can be attributed to the OCP.

Emran, Jiang and Shilpi (2020) also investigate the intergenerational income mobility in China from a demographic perspective, though focusing on gender bias. Based on Becker and Tomes (1986)'s intergenerational mobility model, they incorporate gender bias in the transmission of both human and non-human capital from parents to children. Gender bias against girls in education is observed in urban China. However, girls in China benefit more from their father's higher education than girls in India. Girls with fathers who have more than 14 years of schooling achieve better educational outcomes than boys.

Guo, Xia, and Zhang (2022) further extend the family structure examination to include siblings and parents-in-law in analyzing intergenerational mobility. They apply the Shapley-Shubik-Becker model to examine co-residence competition among multi-child families and between parents and parents-in-law. Factors affecting co-residence decisions include gender, birth order, education, parental education, parent-child age gaps, and local housing prices.

### 4.4. Migration

One prominent feature in contemporary China is its large-scale internal migration accompanying the market reform from 1978. Its ramifications extend to changing family structure and the new dynamics of intergenerational mobility. Shortly after the foundation of the People's Republic of China in 1949, the hukou system (household registration system) was established, which officially classifies individuals as either rural/agricultural or urban/non-agricultural hukou status, at birth cities. Mobility from rural to urban areas within a city or across cities is restricted, especially in the 1960s-1970s. Along with the market reform in 1978, China gradually relaxed the mobility restrictions induced by the hukou system. This shift leads to substantial domestic migration: by the early1990s, approximately 150 million rural Chinese had migrated to urban areas in or outside home cities (Freeman, 2006). The total number of migrant workers reached 252.78 million in 2011, according to the China's National Bureau of Statistics. In 2020, the number of migrant workers further rose to 286 million, accounting for 20% of the country's total population (Guo, Zhang, and Zhou, 2024).

The dynamics of transmission of socioeconomic status across generations also change. Migrant parents, usually moving from less to more developed regions, can provide more financial support to invest in the human capital and healthcare of their children, promoting intergenerational upward mobility. Nevertheless, due to the restricted access to key resources at the destination cities (e.g., education, employment, and healthcare), the migrant parents usually leave their children behind in home cities. The absence of parents may create social and child development issues, deterring upward intergenerational mobility.

The literature demonstrates mixed findings. Cong and Silverstein (2011) show that comparing to non-migrant sons, migrant sons in rural China provide more financial support to their parents, who in return offer childcare for their grandchildren—resulting in a more mutually beneficial exchange. Sun, Huang, and Hong (2012) find that the intergenerational income elasticity of the migrant workers is less than half of the non-migrants. In other words, migration—when combined with human capital investment—significantly enhances upward mobility relative to their parents' generation and helps children from the poor families escape the intergenerational poverty trap.

One by-product of migration is prolonged geographical distance with parents/children and less intergenerational contact. Gruijters (2017) estimates an incidence rate ratio of 0.31 lower of visit frequency for children living in different communities or regions than those residing in the same community or village in China. This restriction due to distance was not compensated by other types of contact. Lei and Chae (2024) support Gruijters's findings on intergenerational contact and highlight that parental absence during childhood is negatively associated with closeness to and frequency of visits to parents in adulthood. A one-tenth standard deviation reduction in the father-child relationship is observed with a prolonged period of separation between father and child. Feng and He (2022) echo the findings by showing that parental rural-to-urban migration significantly and negatively affects children's upward intergenerational mobility by 3.34 percentage points. The adverse impact of parental migration on offspring mainly comes through reduced enrollment in higher education and job opportunities. They also highlight that parental migration is positively associated with the likelihood of offspring migrating for work, implying an intergenerational correlation in migration decisions.

### 4.5. Housing price

Along with China's market reform, the housing price fluctuates with market sentiments rather than being underpriced in the regime of planned economy. As the housing becomes more unaffordable, co-residence with parents becomes an increasingly popular choice among young adults to accumulate savings. Such co-living decision also alters China's intergenerational mobility pattern.

Yuan and Lin (2013) estimate that housing assets explain for 37.85% and 15.47% of intergenerational income mobility in urban and rural China in 2005 respectively, which is larger than the contribution from human capital and social capital. Li and Wu (2019) identify the causal positive impact of rising housing price on intergenerational co-residence of young adults in urban China. Such effect is larger for those without a house, having a lower level of wealth, and relatively younger. Rosenzweig and Zhang (2019) further extend the empirical studies on intergenerational co-residence by incorporating individual savings information. They find that rising housing prices significantly increase intergenerational co-residence. The savings rates of the young relative to the middle-aged conditional on income are also elevated, partially due to the subsidies to the young from sharing housing with parents. It is calculated that Chinese youth's savings rates would be 21% lower if housing prices were at the same ratio to disposable incomes as that in the United States. Guo, Xia, and Zhang (2022) distinguish co-residence of adult children with parents and parents-in-law and use counterfactual experiments to quantify the effects of changes in housing prices on intergenerational co-residence. A 20% rise in housing prices increases the likelihood of co-residence for adult children and parents by 3.5%. It is particularly responsive for female, first-born, and highly educated children and highly education parents.

In sum, the escalating housing prices during China's marketization process significantly increases the intergenerational co-residence rates. Parents who can help with the co-living arrangement contribute indirectly to adult children's saving accumulation; thus the intergenerational socioeconomic status persists. Such channel via housing prices plays an increasingly important role in the intergenerational income transmission.

### 4.6. Market transition and trade liberalization

Empirical evidence shows that the intergenerational socioeconomic mobility declines along with China's transition into market economy (Zhou and Xie, 2019; Fan, Yi, and Zhang, 2021). There is a long theoretical discussion on the implications of political and economic institutions on social mobility and stratification (Parkin, 1971; Giddens, 1973). In China's post-revolutionary era, the emergence of markets provides opportunities for the privileged classes to convert political power into market resources, incentivizing the richer parents to pass on socioeconomic advantages to their children, along with the abolition of egalitarian educational policies (Zhou and Xie, 2019). During China's rapid transition into market-oriented economy, it is found that the rising (declining) intergenerational economic persistence (mobility) is positively associated with (a) increasing share of private enterprises; (b) public expenditure on per capita education, science, culture, and public health; and (c) university enrollment rates (Fan, Yi, and Zhang, 2021). Note that the correlation does not imply causality in this context.

After China's entry into the World Trade Organization (WTO) in 2001, its trade liberalization accelerated, together with a rise in intergenerational mobility. Lou and Li (2022) use the export shocks during China's export expansion period 2000-2007 at prefecture level as a quasi-natural experiment, showing a plausibly causal impact on rising intergenerational education mobility. A leading channel lies in an income effect, which relaxes parents' financial constraints in investment in children's human capital. Similarly, using China's entry into WTO as a quasi-experiment and examining at the individual level, Yu, Gong, and Yi (2025) show that the economic inequality across generations is amplified in Chinese villages. A major channel is through relaxed rural-to-urban migration which leads to considerable income benefits. Children from wealthier, rather than poorer families, are better at catching the migration opportunity along with the trade liberalization.

### 4.7. Psychology and belief

Recent studies also uncover a new mindset channel in intergenerational transmission. Roland and Yang (2019) use China's Cultural Revolution as a quasi-natural experiment and examine the impact of parents' mindsets on children's educational attainment. They find that although high-school graduates from the Cultural Revolution period invest more in their children's human capital, they are less likely to support the idea that effort pays off. Nevertheless, they transmit less of their changed beliefs to the next generation but investing more in the children's education, attempting to safeguard their offsprings from having the same misfortunes.

Huang, Song, and Xie (2023) develop a dual-pathway intergenerational transmission model over multiple generations, identifying two intertwined pathways: a socioeconomic pathway (measured by educational attainment) and a psychological pathway (measured by mindset). They find that the mindset factor is strongly associated with various outcomes, such as personalities, attitudes, and cognitive skills. This set of literature offers new insights on the psychological factors, contributing to a better understanding of intergenerational mobility in China.

## 5. Policy Discussion and Conclusion

In this paper, we review studies in the intergenerational mobility in historical and contemporary China. Similar to other developing countries, estimating the degree of intergenerational economic persistence/mobility in China is challenging, because of a short of administrative data and the selection bias from conventional survey data, as well as lifecycle bias and attenuation bias. Despite data and measurement challenges, scholars endeavour to generate estimates on China's intergenerational mobility in socioeconomic status, as measured by income, education, social class, wealth, and health, from high-quality longitudinal surveys. Taking the most popular measure of socioeconomic status, income and education, as examples, the estimated intergenerational income correlation in market-reform era of China is around 0.4 from multiple studies (Deng, Gustafsson, and Li, 2013; Chen, 2016; Jin et al., 2019; Fan, Yi, and Zhang, 2021), approaching the level of the United States (Chetty et al., 2014). Most intergenerational educational correlation estimates vary between 0.2 and 0.4, depending on gender, cohort, and parental SES (Magnani & Zhu, 2015; Chen et al., 2019; Xie et al., 2022). There is evident decline in intergenerational mobility in China in recent decades and younger cohorts.

Considering channels through which parents' SES is passed down to the next generation in China, most literature shows that the leading factor remains education (Qin, Wang, and Zhuang, 2016; Liu and Wan, 2019; Duan et al., 2022), consistently with the classical human capital theories (Becker and Tomes, 1979, 1986) and China's historical meritocratic educational system (Chen, Kung, and Ma, 2020). Health, as another component of human capital, is also tested as one important channel in intergenerational transmission of SES (Kong, Osberg, and Zhou, 2019), together with social capital, measured by occupation, industry and ownership of the work unit, and party affiliation (Zhang and Eriksson, 2010, Yuan and Chen, 2013; Fan, 2016; Emran et al., 2023).

Because of the special population policies in China, such as the One-Child Policy and the rural vs. urban hukou system, demographic factors play an important role in the intergenerational transmission of SES. On one hand, fertility decision, co-residence decision, and demographic structure matter, as urban wealthier parents are more constrained to have one child only and are incentivized to pass on their wealth to the single child, declining intergenerational mobility (Yu, Fan, and Yi, 2025). On the other hand, gender bias, birth order, parent/child education, parent-child age gaps, and local housing prices, all play important roles in determining co-residence decisions in multi-child families and thus affecting intergenerational mobility patterns (Emran, Jiang, and Shilpi, 2020; Guo, Xia, and Zhang, 2022). With the relaxation in *hukou* system after the market reform in 1978, there is large-scale internal migration from rural to urban areas. It goes with changes in upward mobility across generations, though the findings are mixed, as migration parents are able to accumulate more financial resources to invest in their children's human capital though trade-off with less accompanying time (Cong and Silverstein, 2011; Sun, Huang, and Hong, 2012; Feng and He, 2022; Lei and Chae, 2024).

Along with the transition from planned into market-oriented economy, there is evident increase in the share of private enterprises, escalating housing price, and improving trade liberalization. Such market factors are found associated with changes in the patterns of intergenerational mobility, as parents who benefit from the housing wealth rise, development of private sectors, and trade liberalization are now incentivized to pass down the wealth to the next generation (Yuan and Lin, 2013; Zhou and Xie, 2019; Guo, Xia, and Zhang, 2022; Lou and Li, 2022).

Last but not least, recent literature links intergenerational mobility in China with belief, showing that in addition to the conventional socioeconomic pathway, the psychological pathway measured by mindset, also contributes significantly to the transmission of SES across generations.

To promote intergenerational mobility in China, especially the upward mobility for children born to less advantaged parents, policy efforts are suggested along the following directions. On education, it is advised to improve equal access to human capital for age eligible kids from different family background, to make sure that children of migrant workers and rural residents can access 9-year compulsory education and the scholarship in colleges and universities can target students in need from less favoured families. Second, from a demographic perspective, with relaxation of the OCP and hukou restrictions, it is necessary to develop the professional market for childcare, as well as providing educational and mental support for the left-behind children. Third, with the market reform and privatization process, a shift from connection to merit is recommended, to disincentivize benefited parents to pass down social status or wealth through hidden connections but not merits. Future works are warranted when more administrative data become available to explore new channels in the intergenerational transmission of SES in China.

# Appendix

Table A1. Summary of Mechanisms for Intergenerational Mobility in China

| Panel A Human capital | | | |
|---|---|---|---|
| **Study** | **Data source** | **Time period / Cohort** | **Preferred mechanisms** |
| Chen et al. (2015) | Chinese Urban House-<br><br>hold Education and Employment Survey (UHS) 2004 | 1930-1985 birth cohorts in urban areas | Intergenerational educational mobility follows a U-shaped pattern, with lower rank-rank education persistence in the Maoist era (0.146-0.196), higher during the Pre-Mao era (0.214-0.249) and post-Mao era (0.160-0.393); Transmission difference of fathers' educational status for Maoist and post-Mao era is 0.043-0.077; Mothers' transmission difference is 0.032 |
| Li and Zhou (2015) | CHARLS 2011 & CGSS 2008 | Parent cohorts that aged 45 and above | Intergenerational educational elasticity increases by 0.25 if parents fund their child's college education; Government funding for education reduces the intergenerational elasticity of education by 0.04 |
| Qin, Wang, and Zhuang (2016) | CHNS 1989 to 2009 | Children above 25 years old and has income | IGE of 0.481 after accounting for the direct transfer of human capital; Intergenerational transmission of education is 0.310; Intergenerational transmission of health is 0.485 |
| Yang and Qiu (2016) | CHIP 2002 & 2007 and China Statistics Yearbook 1991 to 1996 | Males between 46 to 60 and 30 to 45 | Intergenerational correlation of 0.48 was found, with 48.4% of the correlation attributed to innate ability; innate ability gap between families in the top and bottom income quantiles is 1.36, and this gap expands to 2.35 after compulsory education |
| Liu and Wan (2019) | CHIP 2007 & 2013 | Students sat for the 1984 to 2010 college entrance examination | Children with high school-educated parents have an average of 0.626 more years of schooling compared to those with uneducated parents; Inverted U-shaped between intergenerational transmission of education and supply of higher education |
| Luo and Liu, (2020) | CHIP 2013 | Cohorts born between 1940 to 1989 | Upward intergenerational educational mobility rate is 0.2902 for father-child and 0.2444 for mother-child, with educational expansion increasing these rates by 0.1416 and 0.345 respectively |
| Huang, Huang, and Shui (2021) | CHNS 1989 to 2015 | Graduated children born between 1973 and 1997, with fathers younger than 65 years old | A high IGE of 0.340 was found in low-spending provinces when total government spending was divided into three groups; 0.296 for medium-spending provinces; 0.177 for high-spending provinces; marginal probability of upward mobility for urban residents is 12.33% and rural residents is 6.46% |
| Duan et al. (2022) | China General Social Survey (CGSS) 2008, 2010, 2011, 2012, 2013, 2015, and 2017 | Cohorts between the ages of 17 and 60 who had completed formal education | A positive coefficient of 0.010 (0.011) for intergenerational persistence in urban (rural) occupational socioeconomic status was found after the expansion of higher education in China (CHEE); CHEE shifted the distribution of higher education opportunities from a quasi-inverted U-shape to a quasi-linear shape |
| Zhang and Gu (2025) | CFPS 2010 - 2018 | Child is defined as an individual aged between 6 and 21 who is currently attending school | Expanding college capacity from 6% to 35% increases intergenerational income persistence by 6-8% |

| **Panel B Social capital** | | | |
|---|---|---|---|
| **Study** | **Data source** | **Time period / Cohort** | **Preferred mechanisms** |
| Zhang and Eriksson (2010) | CHNS 1989 to 2006 | Child aged between 20 to 50 | IGE of 0.45, with 12% of intergenerational persistence accounted for by living in a coastal province and 14% accounted for by being born in an urban area; 23% Gini of inequality of opportunity is accounted by parental household income; 20% by Mother's type of employer; 19% by father's |
| Yuan and Lin (2013) | CHIP 1988 to 2002, Chinese General Social Survey (CGSS) | Individuals aged 20 to 65, with less than a 10-year age difference between parent and child | 60% of intergenerational income mobility is accounted by human capital, social capital and wealth; IGE is accounted for by social capital |
| | 2006 & CHNS 1989 to 2009 | | of 4.73%–9% in urban areas and 6.17%–9.71% in rural areas during 1995–2005 period |
| Fan (2016) | CHIP 1995 to 2002 | Father cohorts aged under 60 and child cohorts that are at least 23 years old and live with parents in urban areas | IGE varies from 0.433 to 0.512 comparing the before and after market reform in China, with a high IGE of 0.71 in post-reform era; Parental investment in social capital contributes significantly to intergenerational income transmission for richer families, with a magnitude of 0.651, while poorer families have a magnitude of 0.109 |
| Kong, Osberg, and Zhou (2019) | CHNS 1997, 2000 and 2004 | Children between 4-18 years old | 1.8 kg gain and 2.2 percentage-point increase in the overweight rate of medium-build 10-year-old boys |
| Emran et al. (2023) | CFPS 2010 | Cohorts aged between 29 to 65 | Intergenerational regression coefficient (IGRC) of 0.311 is found for children whose fathers have farm occupations, and 0.316 for those whose fathers have non-farm occupations |

| **Panel C Fertility and demographic structure** | | | |
|---|---|---|---|
| **Study** | **Data source** | **Time period / Cohort** | **Preferred mechanisms** |
| Emran, Jiang, Shilpi (2020) | CFPS 2010 & 2016 | Children of the 18-35 age cohorts in 2016 | Intergenerational regression coefficient (IGRC) is 0.27 for sons, compared to 0.33 for daughters |
| Guo, Xia, and Zhang (2022) | CFPS 2010 | Urban parents aged between 55 to 75 with children been married | Female adult children's co-residence probability (4.2%) increases by 1% compared to males (3.2%); co-residence probability is 19.9% for children with siblings and 21.0% for those without siblings |
| Yu, Fan, and Yi (2024) | CFPS 2011 to 2015 and CHARLS 2011 to 2015 | Individuals aged 45 years and above | Intergenerational income mobility of 0.26 for 1964 to 1973 cohorts; 0.37 for 1983 to 1985 cohorts; One-child policy accounts for 35.4%-51.5% of China's decrease in intergenerational mobility |

| Panel D Migration | | | |
|---|---|---|---|
| **Study** | **Data source** | **Time period / Cohort** | **Preferred mechanisms** |
| Cong and Silverstein (2011) | Collected survey from 2001 and 2003 waves | Adults aged 60 and above and live in Chaohu region | Adult migrant sons provide more financial support to their older parents than non-migrants, with coefficient between 0.48 to 0.60 |
| Sun, Huang, and Hong (2012) | CGSS 2006 & China statistical yearbook for regional economy 2006 | Adults with yearly income | Migration significantly influence intergenerational income mobility by 0.31 for rural areas; 0.44 for families with high income; 0.51 for those with non-college-educated fathers |
| Gruijters (2017) | CHARLS 2011 to 2012 | Parent cohorts aged 45 and above with non-coresident children | Rate of change in parent-child contact varies between 0.03 and 0.01 for visits among those staying in a different province and those abroad; between 0.63 and 0.48 for other forms of contact |
| Feng and He (2022) | CHNS 1991 to 2015 | Employed individuals experience parental migration from rural to urban areas at child age of 0 to 18 | Exposure to parental migration and being left-behind children reduces intergenerational upward mobility by 3.34 percentage points and increases downward mobility by 3.90 percentage points |
| Lei and Chae (2024) | CFPS 2010 to 2016 | Adults between 25 and 54 years old who have at least one living parent in 2016 | Separation from the mother during childhood reduces parent-child closeness by 0.110 units and frequency of visits by 0.183 units; separation from the father reduces parent-child closeness by 0.066 units and frequency of visits by 0.218 units |

| Panel E Housing price | | | |
|---|---|---|---|
| **Study** | **Data source** | **Time period / Cohort** | **Preferred mechanisms** |
| Li and Wu (2019) | CHNS 1991 to 2011 and Inter-Census Population Survey 2005 | Married adults between 20 to 50 years old in urban areas and have at least one parent alive | Rise in housing prices accounts for 0.317 percentage-point of intergenerational co-residence with elderly parents or parents-in-law; homeownership reduces intergenerational co-residence probability by 2 percentage-point |
| Rosenzweig and Zhang (2019) | Chinese Twin Survey, Chinese Non-Twin Survey, CHIP 2002 & 2013 | Young adults between 25 to 34 years old & Parents aged between 45 to 64 | 14% increase in co-residence probability between parents and adult children is accounted for by quality-adjusted housing prices |
| Guo, Xia, and Zhang (2022) | CFPS 2010 | Urban parents aged between 55 to 75 with children been married | Housing price rise results to a 0.6 percentage-point (3.2%) of co-residence probability among parent and adult children |

| **Panel F Market transition and trade liberalization** | | | |
|---|---|---|---|
| **Study** | **Data source** | **Time period / Cohort** | **Preferred mechanisms** |
| Zhou and Xie (2019) | Life Histories and Social Change in Contemporary China (LHSCCC) 1996 & CGSS 2005, 2006, 2008, 2010 and 2012 | Cohorts aged 31 to 64 and born between 1936 and 1981 | Effect of socioeconomic status hierarchy increased considerably for cohorts born between 1950 and 1960; Significant curvilinear trend in intergenerational immobility among the farming class |
| Fan, Yi, Zhang (2021) | CFPS 2010 to 2016 | 1970-1988 birth cohorts | An increase of 0.201 IGE for coastal regions was found among the 1970–1980 and 1981–1988 cohorts while the change in IGE for inland regions is -0.012; 0.104 for urban areas; 0.049 for rural areas; Overall increase of 0.052 IGE among the cohorts |
| Luo and Li (2022) | CFPS 2010 | Child cohort aged between 20 to 24 in 2010 | A 0.9741 year increase in education from export shock was found to impact intergenerational educational mobility for children with parents of education levels below the 25th percentile; Overall 0.3206 years increase on education mobility from export shock |
| Yu, Gong, and Yi (2025) | CHIP 1995, 2002, 2007, 2013, Research Center for the Rural Economy (RCRE) survey 2003 to 2013, World Bank Trade | Son cohorts born between 1966-1981 and 1982-1994, and have parents who | Income rank-rank slope increases from 0.38 for the cohort born between 1966 and 1981 to 0.55 for the cohort born between 1982 |
| | Analysis and Information System (TRAINS) 2013, Chinese Industrial Enterprises database 2000 & 1% sample of 2000 Chinese population census | were at most 64 years old in 2013, with the age difference between them ranging from -5 to 12 years | and 1994; Trade shocks increase the rank-rank slope by 0.25 as exposure rises from the 25th to the 75th percentile |

| **Panel G Psychology and belief** | | | |
|---|---|---|---|
| **Study** | **Data source** | **Time period / Cohort** | **Preferred mechanisms** |
| Roland and Yang (2017) | CFPS 2010 & 2012 | Cohorts born between 1957 and 1960 who have completed at least a high school education | Intergenerational elasticity of belief ranges between 0.283 to 0.363 for belief in hard work and between 0.241 to 0.303 for belief in local government |
| Huang, Song, and Xie (2023) | CFPS 2010 to 2018 | Child cohorts aged 12 and above in 2010, aged 10 to 15 in 2012 and 2014; Parents cohorts born between 1940 and 1980; | Intergenerational educational mobility (rank-rank correlation) of children from growth mindset families is 0.509; 0.579 for those with fixed mindset parents; Intergenerational mindset mobility (rank-rank correlation) is 0.25 from parent to child; 0.281 for high education parents & 0.211 for low education parents |